\pdfoutput=1
\documentclass{sig-alternate}

\usepackage{moreverb}
\usepackage[square,sort,comma,numbers]{natbib}
\usepackage{amsmath,amssymb, mathtools}
\newtheorem{theorem}{Theorem}

\newenvironment{definition}[1][Definition]{\begin{trivlist}
		\item[\hskip \labelsep {\bfseries #1}]}{\end{trivlist}}

\usepackage{graphicx}
\graphicspath{ {./} }

\newcommand\BibTeX{{\rmfamily B\kern-.05em \textsc{i\kern-.025em b}\kern-.08em
T\kern-.1667em\lower.7ex\hbox{E}\kern-.125emX}}

\newcommand{\xhdr}[1]{\vspace{1.7mm}\noindent{{\bf #1.}}}

\makeatletter
\renewcommand{\@ptsize}{0}
\makeatother
\usepackage{setspace}

\begin{document}


\title{A two-stage working model strategy for network analysis under Hierarchical Exponential Random Graph Models}

\numberofauthors{1}
\author{
	\alignauthor
	Ming Cao\\
	\affaddr{University of Texas Health Science Center at Houston}\\
	\email{ming.cao@uth.tmc.edu}
}

\maketitle

\begin{abstract}
Social networks as a representation of relational data, often possess multiple types of dependency structures at the same time. There could be clustering (beyond homophily) at a macro level as well as transitivity (a friend's friend is more likely to be also a friend) at a micro level. Motivated by \cite{schweinberger2015local} which constructed a family of Exponential Random Graph Models (ERGM) with local dependence assumption, we argue that this kind of hierarchical models has potential to better fit real networks. To tackle the non-scalable estimation problem, the cost paid for modeling power, we propose a two-stage working model strategy that first utilize Latent Space Models (LSM) for their strength on clustering, and then further tune ERGM to archive goodness of fit.
\end{abstract}


\footnotetext[2]{}

\section{Introduction}
\begin{figure*}
	\centering
	\begin{minipage}[b]{0.4\textwidth}
		\includegraphics[width=\textwidth]{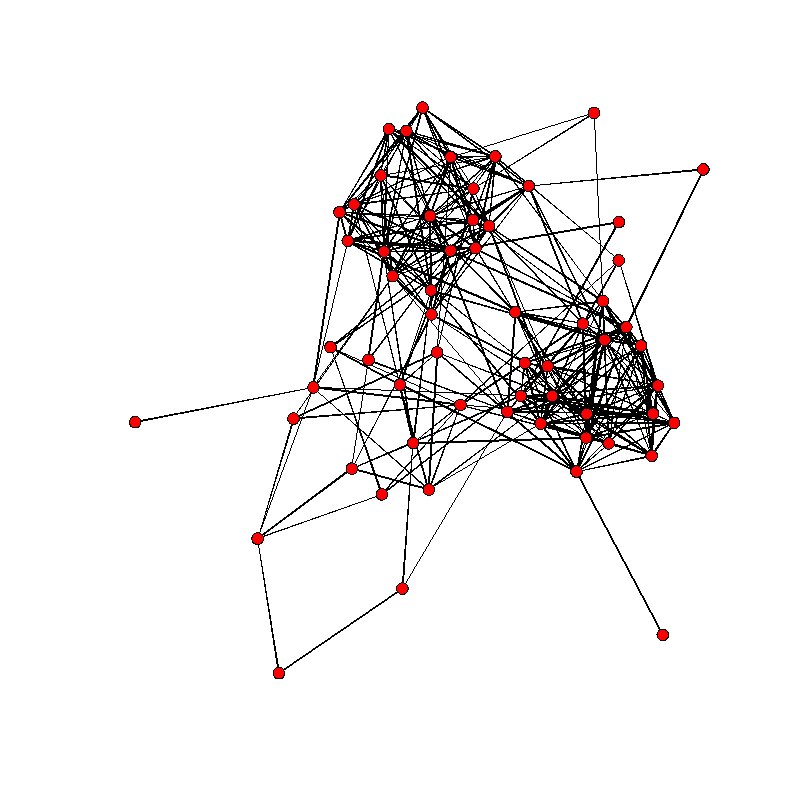}
	\end{minipage}
	\hfill
	\begin{minipage}[b]{0.4\textwidth}
		\includegraphics[width=\textwidth]{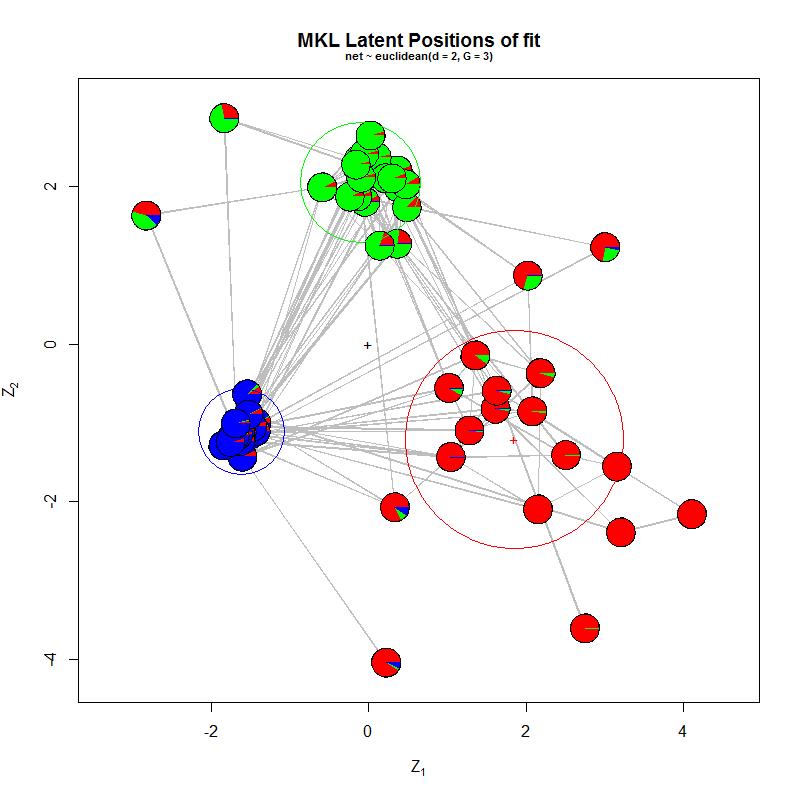}
	\end{minipage}
	\caption{A simulated network to illustrate its hierarchical structure: on the left is a mix of three communities each has 20 nodes, between-cluster ties $Y_{i,j}$ ($i\in k \text{th cluster}$, $j\in l \text{th cluster}$, $k \ne l$) are i.i.d. Bernoulli(0.05). The clusters are generated from an exponential random graph model with a baseline density of $0.05$ plus a transitivity parameter (in terms of gwdsp and gwesp, which will be explained in Section \ref{ergm}) of $0.2$, $0.5$ and $1.0$, respectively. The cluster of low transitivity is not visually apparent. On the right is the clustered visualization of this network by a latent space model (see Section \ref{lsm}).}
	\label{fig_demo}
\end{figure*}

Social networks take the form of a graph consisting a set of nodes and edges. Typically, the nodes represent persons or organizations, and each edge is a measure of the relation between a pair of nodes. For example, in the citation network of Statisticians \cite{ji2014coauthorship}, a (directed) link variable $Y_{i,j}$ indicates individual $i$ has cited $j$'s work if $Y_{i,j} = 1$, otherwise if $Y_{i,j} = 0$. Statistical analysis beyond descriptive is focused on modeling dependencies of the link formation.\\
As data getting collected at larger scales, social networks often exhibits a hierarchical structure: people are from different communities, within each community there are various types of link formation process taking action while between communities the connections are much sparser. Communities could be not only physical such as geographical, but also abstract such as by political attitude. Within a community, transitivity is often a major type of force that generate links, for instance, if $i$ cites $j$ and $j$ cites $k$, then it is more likely that $i$ also cites $k$. However, we should also expect different strength of transitivity among statisticians working in different areas (clustering). Figure \ref{fig_demo} is a simulated (undirected) network of three communities each has the same number of nodes (20) but with different transitivities, while any probability of a between-cluster tie is the same (0.05). Using the R package \textit{latentnet} \citep{krivitsky2008latentnet}, we can see the structures clearly: the cluster with high transitivity (blue) has its nodes closer to each other, and the one with low transitivity (red) is more spreading. The uncertainties of clustering are indicated by the pie of colors.\\
When the network is homogeneous, i.e. has one single community, Exponential Random Graph Models (ERGM) is a popular tool for modeling as it provides researchers an intuitive formulation of related structures to test various social theories \citep{wasserman1996logit}. The Monte Caolo Markov Chain (MCMC) techniques developed in \cite{snijders2002markov,hunter2006curved,hummel2012improving} and computer programs \textit{statnet} \citep{handcock2008statnet} led to its widespread use. However, practitioners often found that the programs had convergence problems for many specifications, and the statistical properties of the MLE are not comprehensive. Recently, \cite{schweinberger2015local} suggests that the distribution of sufficient statistics in the traditional ERGM can be asymptotically normal if some local dependence is imposed. Hence solves the notorious degeneracy problem, which is mainly caused by the global dependence introduced by the Markov property \citep{frank1986markov}. However, the Bayesian inference procedure proposed by their paper is extremely expensive in computation as it involves two exponentially increasing terms, one nested in the other.\\
\xhdr{Present work}
Motivated by the Hierarchical ERGM construction in \cite{schweinberger2015local}, we attempt to find a general and feasible procedure to infer clustering and the within-cluster dependences simultaneousy. However, our purpose is not to archive the large sample properties as the number of clusters goes to $\infty$ in a Frequentist way, because we view the network as a fixed set of nodes and the parameter estimates not only for interpretation but also as a parsimonious local mechanism to represent global structure, e.g. improving model goodness of fit \citep{hunter2012goodness}. \\
To illustrate our ideas more efficiently, from here we focus on the \textit{undirected binary} graph  though the extension to the \textit{directed} and/or \textit{weighted} graph is straightforward. Notations are given as following: an \textit{undirected binary graph} $G=(V,E)$ consists a set of \textit{vertices} $V=\{1, \dots, n\}$ and \textit{edges} $E=\{(i,j) | i,j \in V \}$. Typically, $G$ is represented by its adjacency matrix $Y = \{Y_{i,j}\}_{1\le i\ne j\le n}$ where
\[
Y_{i,j}=\begin{cases}
1 & \text{if there is an edge between vertices } i \text{ and } j\\
0 & \text{otherwise}
\end{cases}
\]
and $Y_{i,i}=0$ for all $i$ as loops are not allowed. We may also observe pair-specific characteristics $X=\{x_{i,j}\}$. It could be a function of node-specific attributes, for example, $x_{i,j}=I(x_i = x_j)$ where $I$ is the indicator function. This so called \textit{homophily} means individuals with similar attributes are more(less) likely to be linked. Furthermore, the \textit{vertices} belong to $K$ clusters/neighborhoods/blocks, so each of them has membership/color $m_i=k$ for $k \in {1, \dots, K}$. In math literatures, $(Y, M=\{m_i\})$ is called \textit{colored graph}.\\
In Section \ref{sec_ergm} we take a brief review of recent works on ERGM with a two folded purpose. One is to argue that with a hierarchical construction, it has capability to serve as a (hypothetical) true model for networks in the real world. The other is to show difficulty in estimation. In Section \ref{sec_TwoPhase}, we uncover its connection to LSM by changing the angle of view, hence propose our two-stage strategy using a working model. In Section \ref{sec_examples}, two examples are given to apply the proposed strategy, which is essentially about how to choose a working model appropriately. We summarize our idea, discuss its limitation, and point out some directions to improve in Section \ref{sec_disc}. 

\section{The Evolution of ERGM} \label{sec_ergm}
\cite{frank1986markov} defined a probability distribution for a graph to be a Markov graph, if the number of nodes is fixed at $n$ and possible edges between disjoint pairs of nodes are independent conditional on the rest of the graph. It is motivated by the Markov property of stochastic process and spatial statistics on a lattice \citep{besag1974spatial}. With the Hammersley-Clifford theorem, and under the permutation invariance assumption, it is proved that a random undirected graph is a Markov graph if and only if the probability distribution can be written as
\begin{equation} \label{markov_graph}
P_{\theta}\{Y = y\} = \text{exp}\left( \sum\limits_{k=1}^{n-1}\theta_k S_k(y) + \tau T(y) - \psi(\theta, \tau)\right)
\end{equation}
where the statistics $S_k$ and $T$ are defined by
\begin{align}
S_1(y) &= \sum_{1 \leq i < j \leq n}y_{ij} & \text{number of edges} \nonumber \\
S_k(y) &= \sum_{1 \leq i \leq n}{y_{i+} \choose k} & \text{number of k-stars} (k \geq 2)\\
T(y) &= \sum_{1 \leq i < j < h \leq n}y_{ij}y_{ih}y_{jh} & \text{number of triangles} \nonumber
\end{align}
with $\theta_k$ and $\tau$ denoting the parameters, and $\psi(\theta, \tau)$ is the normalizing constant. A practical model will truncate $k$ to a small number say 2, i.e. the sufficient statistics is a vector of count for how many edges, 2-stars and triangles are in the graph. \cite{wasserman1996logit} further proposed to use a model of this form with arbitrary statistics $S(y)$ in the exponent which yields the probability functions:
\begin{equation} \label{ergm distribution}
	P_{\theta}\{Y = y\} = \text{exp}\left({\theta}'S(y) - \psi(\theta)\right)
\end{equation}
where $S$ can be a vector of any dimension, so that it leaves space for researchers to specify structures of scientific interests. The interpretation of the parameters is typically based on the log odds ratio of forming a tie, conditional on the rest of the graph since:
\begin{equation} \label{conditional}
	\text{logit}\left( P_{\theta}\{ Y_{i,j}=1 | Y^c_{i,j} \} \right) = {\theta}^{'} c_{i,j}
\end{equation}
where $Y^c_{i,j}=\{Y_{u,v} | \text{ for all } u<v, (u,v)\neq(i,j)\}$ represents all other ties except $Y_{i,j}$, $c_{i,j} = S\left(y^{(ij1)}\right) - S\left(y^{(ij0)}\right)$ is the \textit{change statistic} with $y^{(ij0)}$ and $y^{(ij1)}$ denoting the adjacency matrices with the $(i,j)$th element equal to $1$ and $0$ while all others are the same as $y$. One example using formula \ref{markov_graph} is that when the triangle parameter $\tau$ is positive, the log odds of forming a tie will increase by $\tau$ if this tie also completes a triangle (conditional on the status of all other ties in the graph). It is an indication of transitivity, which means that if we have a friend in common, we are more likely to be friends. Not only facilitated a good interpretation, this conditional formula \ref{conditional} also induced a pseudo-likelihood \citep{strauss1990pseudolikelihood} defined by $l(\theta) = \sum\limits_{i<j}ln\left( P_{\theta}\{ Y_{i,j}=y_{i,j} | y^c_{i,j} \} \right)$, which is a sum of the (log) conditional probabilities and can be fitted by a logistic regression.
\subsection{Estimation is difficult}
However, the maximum likelihood estimation has a major barrier that the normalizing function $\psi(\theta) = \text{log}\sum\limits_{y \in \mathcal{Y}} \text{exp} ({\theta}'S(y)) $ is typically intractable. The summation is over the sample space $\mathcal{Y}$ where the number of possible graphs became astronomically large even when the number of nodes are only dozens. To tackle this intractable likelihood problem, \cite{snijders2002markov} proposed an Markov Chain Monty Carlo (MCMC) way for approximating the ML by following the approach of \cite{geyer1992constrained}. Random samples from the distribution \ref{ergm distribution} can be obtained using the Gibbs sampler \citep{geman1984gibbs}: cycling through the set of all random variables $Y_{i,j}$ ($i \neq j$), or by mixing \citep{tierney1994mixing}, to generate each value according to the conditional distribution in \ref{conditional}. A comparison of the statistical properties of the Maximum Likelihood Estimator (MLE) and Maximum Pseudo Likelihood Estimator (MPLE) showed that MLE could perform much better than MPLE on Bias, Efficiency and Coverage Percentage, especially in terms of the mean value re-parametrization $\mu(\theta) = {\textbf{E}}_{\theta} \left[ S(Y)\right]$ \citep{van2007comparison}.\\
A problem of the above mentioned sampling approximation to the ML inference of ERGM, termed \textit{inferential degeneracy}, persists as an obstacle to real applications. While it appears to be a MCMC algorithm issue of not converging or always converging to a degenerate (complete or empty) graph, this problem is also rooted in the geometry of Markov Graph Models as an exponential family \citep{handcock2003assessing}. There are two lines of efforts to fix it, one line along \cite{snijders2006new} is introduced here and adopted in our proposed procedure, the other line initiated by \cite{schweinberger2015local}, which motivated this paper, will be detailed in the rest of this section. \cite{snijders2006new} extended the scope of modeling social networks using ERGM by representing transitivity not only by the number of transitive triads, but in other ways that are in accordance with the concept of partial conditional independence of \cite{pattison2002neighborhood}. This type of dependence formulates a condition that takes into account not only which nodes are being potentially linked, but also the other ties that exist in the graph: i.e., the dependence model is realization-dependent. Specifically, it states that two possible edges with four distince nodes are conditionally dependent whenever their existence in the graph would create a four-cycle. Along this line, \cite{hunter2006curved} proposed the Curved Exponential Family Models and \cite{hummel2012improving} proposed a lognormal approximation and "stepping" algorithm. Together with the development of a suite of R packages called \textit{statnet}, the applied work began to adopt the ML inference widely.
\subsection{Hierarchical ERGM}
Finally it comes to the model exactly motivated our work. Inspired by the notion of finite neighborhoods in spatial statistics and M-dependence in time series, \cite{schweinberger2015local} proposed the local dependence in random graph models, which could be constructed from observed or unobserved neighborhood structure. Their paper shows that while the conventional ERGM do not satisfy a natural domain consistency condition, the local dependence satisfy it such that a central limit theorem can be established. Their effort is trying to fix the fundamental flaw of Markov random graph models that, for any given pair of nodes $\{i,j\}$, the number of neighbors is $2(n-2)$ and thus increases with the number of nodes $n$. This insight leads to a natural and reasonable assumption that each edge variable depends on a finite subset of other edge variables. They define:
\begin{definition}{(local dependence)}
	The dependence induced by a probability measure $\mathbb{P}$ on the sample space $\mathbb{Y}$ is called \textit{local} if there is a partition of the set of nodes A into $K \geq 2$ non-empty finite subsets $A_1, \dots, A_K$, called neighbourhoods, such that the within- and between- neighbourhood subgraphs $Y_{k,l}$ with domains $A_k \times A_l$ and sample spaces $Y_{k,l}$ satisfy,
	\begin{equation}
		P_K(Y \in \mathcal{Y}) = \prod_{k=1}^{K} \mathbb{P}_{k,k}(Y_{k,k} \in \mathcal{Y}_{k,k}) \prod_{l=1}^{k-1}\mathbb{P}_{k,l}(Y_{k,l} \in \mathcal{Y}_{k,l}, Y_{l,k} \in \mathcal{Y}_{l,k})
	\end{equation}
	where within-neighborhood probability measures $\mathbb{P}_{k,k}$ induce dependence within subgraphs $Y_{k,k}$, whereas between-neighborhood probability measures $\mathbb{P}_{k,l}$ induce independence between subgraphs
\end{definition}
Thus, local dependence breaks down the dependence of random graph $Y$ into subgraphs, while leaving Scientists freedom to specify dependence of interest within subgraphs. Under some sparsity condition, local and sparse random graphs tend to be well behaved in the sense that neighborhoods cannot dominate the whole graph and the distribution of statistics tends to be Gaussian, provided the number of neighborhoods K is large.\\
To estimate, they proposed a fully Bayesian approach. With the following conditional likelihood:
\begin{equation}
	P(Y=y | M=m) = \prod_{k=1}^{K}P(Y_{k,k}=y_{k,k} | M=m)\prod_{l=1}^{k-1}P(Y_{k,l}=y_{k,l} | M=m)
\end{equation}
where the between-neighborhood ties are assumed to be independent $P(Y_{k,l}=y_{k,l} | M=m) = \prod_{i \in A_k, j \in A_l}P(Y_{i,j} = y_{i,j} | M=m)$, the within-neighborhood probability has specific ERGM parameters as $P_{\theta}(Y_{k,k}=y_{k,k} | M=m) = exp\{\theta_k'S_k(y_{k,k}) - \psi_k(\theta_k) \}$.
The marginal distribution of membership is assumed as $M_i \overset{\text{iid}}{\sim} \textbf{Multinomial}(\pi)$, for all $i=1, \dots, n$. For the sake of illustration purpose, we omit the non-parametric priors on the neighborhood structure (membership), only stating the parametric one here:
\begin{align*}
\pi = (\pi_1, \dots, \pi_K) & \sim Dirichlet(\omega_1, \dots, \omega_k) \\
\theta_k & \overset{\text{iid}}{\sim} \textbf{MVN}(\mu_k, \Sigma_k^{-1}) & k=1, \dots, K\\
\theta_B & \sim \textbf{MVN}(\mu_B, \Sigma_B^{-1})
\end{align*}
where $\theta_B = \{\theta_{kl} | k \ne l; k,l \in 1,\dots,K\}$ is a vector of parameters governing the between-neighborhood distribution. It could be simplified to just one scalar $p$ by assuming all between-neighborhood ties are i.i.d. Bernoulli($p$) as in Figure \ref{fig1} and Section \ref{sec_TwoPhase}.
 
\section{Two-stage estimation} \label{sec_TwoPhase}
In this section, we change the angle of viewing Hierarchical ERGM (HERGM) to uncover its connection to another widely used class of network models, namely Latent Space Models (LSM). Recall that our major concern is unobserved (latent) clustering structure will ``confound'' the true within-cluster effect(s). If this is a bottom-up way of first pick a specific ERGM and then consider (possibly) multiple communities, now for estimation we could follow a top-down direction of first tackle clustering problem while taking local structures into account.

\subsection{HERGM as a extension of Stochastic Blockmodels}
A careful inspection of the Bayesian formulation of the HERGM reveals its connection to another class of network models which is initially intended for community (block) detection, namely Stochastic Blockmodels (SBM) \citep{snijders1997block,nowicki2001block}. The purpose of blockmodeling is to partition the vertex set into subsets called \textit{blocks} in such a way that the block structure and the pattern of edges between the blocks capture the main structural features of the graph. \cite{lorrain1971block} proposed blockmodeling based on the concept of \textit{structural equivalence}, which states that two vertices are structually equivalent (belong to the same block) if they relate to the other vertices in the same way. The adjacency matrix should show a block pattern if it is permuted in a certain way. So that type of models are formulated this way: 
\begin{definition}{(Stochastic Blockmodels)}
	membership $(M_i)_{i=1}^n$ are assumed i.i.d. random variables with $P(M_i = k) = \pi_k$ for $k=1,\dots, K$ and conditional on $M=\{M_i\}$, the edges $Y_{i,j}$ are independent Bernoulli($\theta_{M_i, M_j}$).
\end{definition}
If we keep the assumptions on membership and between-block edges but relax the independence to ERGM for within-block edges, it became the HERGM essentially. While this extension is conceptually attractive, the computational cost is prohibitive as it involves two exponentially increasing functions. In the SBM part, the sample space of membership $M$ is $K^n$ where $K$ is the number of blocks and $n$ is the number of nodes. In the within-block ERGM part, the sample size of edge variable $Y_k$ is $2^{n_k \choose 2}$ where $n_k$ is the number of nodes in $k$th block (each pair of nodes can have a link present or absent in an undirected binary network). So both parts need MCMC or other sampling methods to do the Bayesian inference or approximate the MLE (see Section \ref{sec_ergm}), directly combining them makes the problem \textit{intractable}. To provide a more feasible way to tackle the inference of HERGM, we import LSM to account for local structures in an indirect way.
\subsection{Generalized to Latent Space Models} \label{lsm}
Instead of explicitly modeling dependence, the Latent Space Models (LSM) postulate latent nodal variables $Z$ and conditional independence of $Y_{i,j}$ given those variables $Z_i$ and $Z_j$. SBM can be viewed as a simple case of LSM in which membership $M_i$ is the only $Z_i$.\\
\cite{hoff2002latent} introduced the concept of unobserved "social space" within which each node has a position so that a tie is independent of all others given the unobserved positions of the pair it connects to:
\begin{equation}
P(Y=y | Z, X, \beta) = \prod_{i \neq j}P(Y_{i,j}=y_{i,j} | x_{i,j}, z_i, z_j, \beta)
\end{equation}
where $X$ are observed covariates, and $\beta$ and $Z$ are parameters and positions to be estimated. \cite{handcock2007cluster} took a subclass Distance Models where the probability of a tie is modeled as a function of some measure of distance between the latent space positions of two nodes:
\begin{equation}
\text{logit}\{P(Y_{i,j} = 1 | x_{i,j}, z_i, z_j, \beta)\} = \beta_0^{T}x_{i,j} - \beta_1|z_i - z_j|
\end{equation}
with restriction of $\sqrt{\frac{1}{n}\sum_{i}{|z_i|}^2} = 1$ for the identification purpose. Then they imposed a finite mixture of multivariate Gaussian distribution for $z_i$ to represent clustering:
\begin{equation}
z_i \overset{\text{iid}}{\sim} \sum\limits_{k=1}^{K} \lambda_k \textbf{MVN}_{d}(\mu_k, \sigma_k^{2}\textbf{I}_d)
\end{equation}
where non-negative $\lambda_k$ is the probability that an individual belongs to the $k$th group, with $\sum\limits_{k=1}^{K}\lambda_k = 1$. A fully Bayesian estimation of this Latent Position Cluster Model was proposed by specifying the priors:
\begin{align*}
\beta & \sim \textbf{MVN}_p(\epsilon, \psi) \\
\lambda & \sim \textbf{Dirichlet}(\nu) \\
\mu_k &\overset{\text{iid}}{\sim} \textbf{MVN}_{d}(0, \omega^2\textbf{I}_d) & k=1, \dots, K\\
\sigma_k^2 &\overset{\text{iid}}{\sim} \sigma_0^{2}\textbf{Inv}\chi_{\alpha}^2 & k=1, \dots, K
\end{align*}
where $\epsilon$, $\psi$, $\nu = (\nu_1, \dots, \nu_G)$, $\sigma_0^{2}$, $\alpha$ and $\omega^2$ are hyper-parameters. The posterior membership probabilities are:
\begin{equation}
P(M_i = k | \text{others}) = \frac{\lambda_k \phi_d(z_i;\mu_k,\sigma_k^2 I_d)}{\sum\limits_{l=1}^{K}\lambda_l \phi_d(z_i;\mu_l,\sigma_l^2 I_d)}, 
\end{equation}
where $\phi_d(\cdot)$ is the \textit{d}-dimensional multivariate normal density.
\subsection{Working Model Strategy}
Now a natural idea comes: can we use LSM as a working model to infer the membership in the HERGM, and then use this information to infer cluster-specific ERGM? Our initial attempt is by \cite{gong1981twostage} type of pseudo maximum likelihood estimation as the following theorem tells us:
\begin{theorem}{(\cite{gong1981twostage})}
	Let $Y_1, \dots, Y_n \overset{\text{iid}}{\sim} F_{\theta_0, \pi_0}$, and let $\hat{\pi}_n(Y_1, \dots, Y_n)$ be a consistent estimate of $\pi_0$. Under certain regularity conditions, for $\epsilon>0$, let $A_n(\epsilon)$ be the event that there exists a root $\hat{\theta}$ of the equation
	\begin{equation}
	\frac{\partial}{\partial \theta} \ell(\theta, \hat{\pi}) = 0
	\end{equation}
	for which $|\hat{\theta} - \theta_0| < \epsilon$. Then, for any $\epsilon>0$, $P\{A_n(\epsilon)\} \rightarrow 1$.
\end{theorem}
when the pseudo maximum likelihood equation has a unique solution, then the pseudo MLE is consistent. The analog in our application is that when the clustering estimator, e.g. the posterior membership predictor of LSM, has good large sample properties, the MLE of cluster specific ERGM parameters conditioned on it should does. However, the problem is that those two models, HERGM and LSM, may be uncongenial to each other, meaning that no model can be compatible with both of them \citep{meng2014trio}. Apparently, they make very different assumptions about data, as in ERGM they follow an exponential family and in LSM they are conditional independent, as well as the ERGM MLE is a frequentist's procedure and LSM is of Bayesian. So is there a way to show they are operationally, although not theoretically, equivalent? In other words, can LSM fully capture the network structures in the true underlying generating mechanism (assumed to be HERGM), to the extent that membership estimator is consistent. \\
\cite{snijders1997block} proposed a property called \textit{the asymptotically correct distinction of vertex colors}, which means that the probability of correctly identifying the membership (color) for all nodes tends to $1$ as $n$ goes to $\infty$. The implication of this property is that once we can find a function $F(Y)$ such that 
\begin{equation} \label{asym_color}
P(M = F(y) | \theta) \rightarrow 1 \qquad \text{for all } \theta \text{ as n} \rightarrow \infty
\end{equation}
then any statistical test or estimator $T(Y,F(Y))$ has asymptotically the same properties as $T(Y, G)$:
\begin{equation}
\lim_{n \rightarrow \infty} P(T(Y,F(Y))=T(Y, M) | \theta) = 1
\end{equation}
Note that $T(Y, G)$ is when membership $G$ is observed but $T(Y,F(Y))$ is based on network $Y$ only. In our situation, the probability $P$ is under HERGM and the function $F$ is through LSM. 
\section{Applications} \label{sec_examples}
In this section, we give two examples of how to apply our working model strategy. From basic ideas in ERGM and LSM, we can see the point that although they impose very different assumptions, their targeting network structures, e.g. homophily, degree heterogeneous, and transitivity, could be the same. Since both of them are a class of models rather than a single model, the key property of an appropriate working LSM model is targeting networks structures as close as possible to the hypothesized true ERGM model.\\
\subsection{A Transitivity Example}
We first specifically consider one important example where the only dependence are within-cluster transitivities. Without generality, we assume between-cluster densities are all equal since the likelihood governed by those nuisance parameters are completely factored out, and the estimates are trivial. The probability mass function is as following:
\begin{eqnarray}
P(Y=y | M=m) &=& p^{y_B}(1-p)^{n_B-y_B} \nonumber\\
& &\prod_{k=1}^{K}\text{exp}\big( \theta_1^{(k)} \text{Edges}(y) + \theta_2^{(k)} \text{GWDSP}(y) \nonumber\\
& & + \theta_3^{(k)} \text{GWDEP}(y) - \psi(\theta^{(k)})\big)
\end{eqnarray}
where the number of between-cluster edges $y_B = \sum\limits_{i \in A_k, j \in A_l}^{k \ne l} y_{i,j}$ follows a Binomial distribution with total number of possible ties $n_B = \sum\limits_{k \ne l} n_k * n_l$ and probability $p$. Each cluster has two statistics, Geometrically Weighted Dyadwise Shared Partner (GWDSP) and Geometrically Weighted Edgewise Shared Partner (GWESP) (see \cite{snijders2006new} for details), to represent the transitivity. Since there are no homophily or degree heterogeneity, we can also omit the covariates and node specific random effects in LSM \citep{krivitsky2009representing} and simply have the probability condition on the distance between latent positions only:
\begin{equation}
\text{logit}\{P(Y_{i,j} = 1 | x_{i,j}, z_i, z_j, \beta)\} = \beta_1|z_i - z_j|
\end{equation}
As long as the defined distance satisfies triangle inequality, it captures the transitivity in the sense that if $i$ and $j$ are both close to $k$, then $i$ and $j$ should be also close to each other. Intuitively, if the true model has the transitivity as its only dependency structure, then the working model should be able to recover the membership.
\subsubsection{Stage 1: clustering}
\begin{figure*}
	\includegraphics[width=\textwidth]{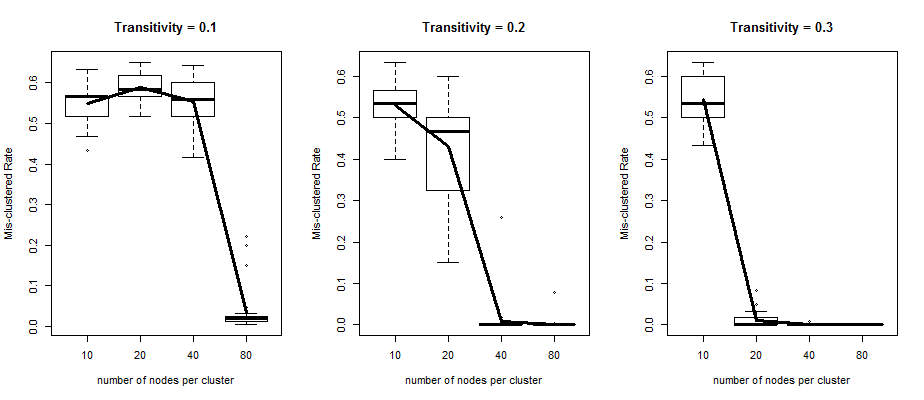}
	\caption{The mis-clustered rates drop as the number of nodes per cluster increase, with different speed at different transitivity strengths.}
	\label{fig_misrate}
\end{figure*}

First we evaluate the performance, in terms of mis-clustering rate, of the working model along sample size and transitivity strength. From Figure \ref{fig_misrate}, we can see that mis-clustering rates drop as sample size increase, and the stronger the transitivity, the faster it hits zero.
\subsubsection{Stage 2: fine tuning}
\begin{figure*}
	\centering
	\begin{minipage}[b]{0.45\textwidth}
		\includegraphics[width=\textwidth]{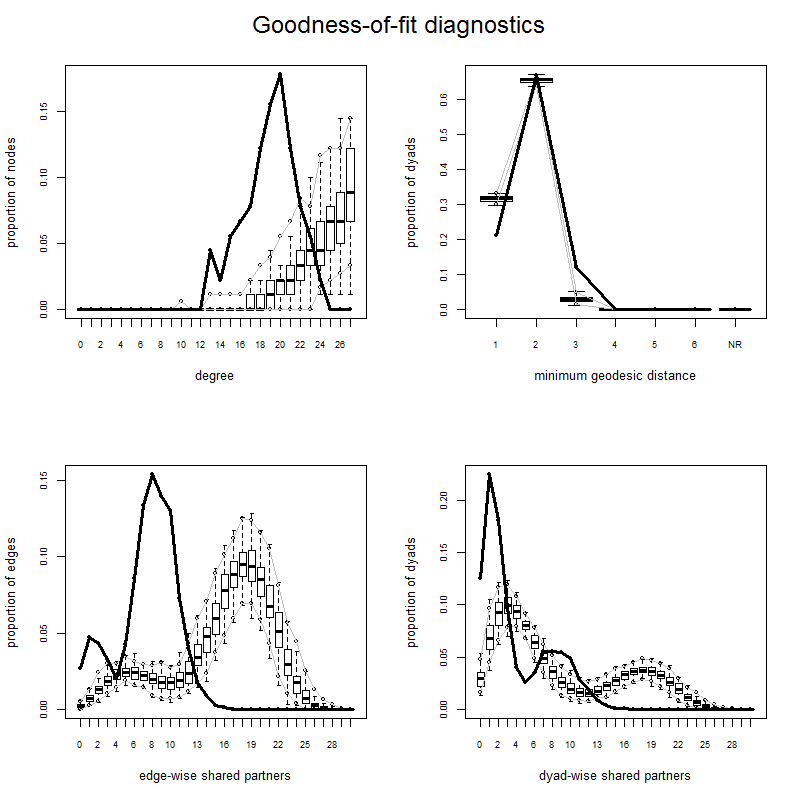}
	\end{minipage}
	\hfill
	\begin{minipage}[b]{0.45\textwidth}
		\includegraphics[width=\textwidth]{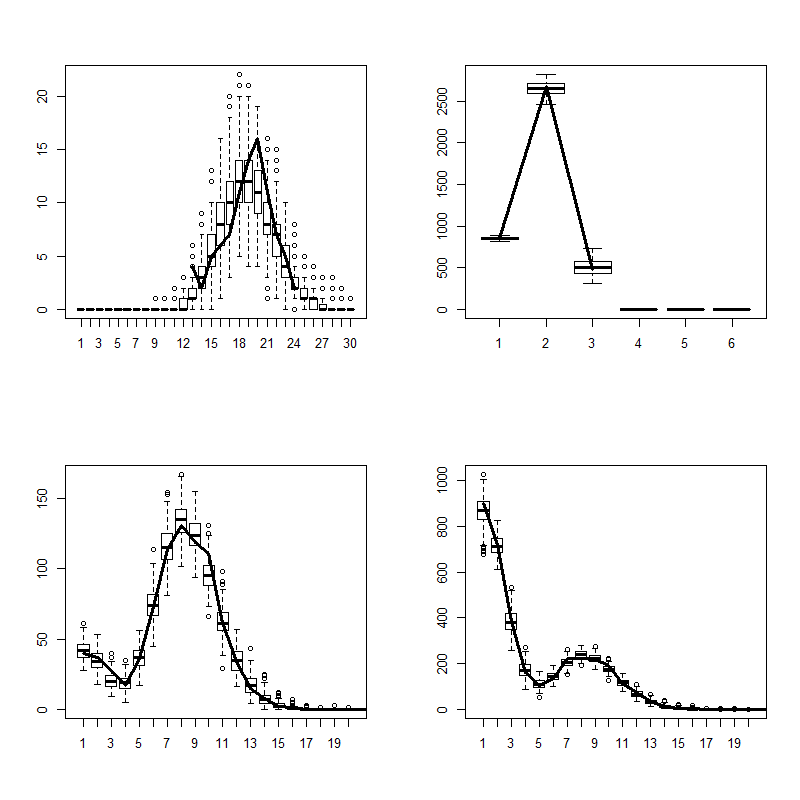}
	\end{minipage}
	\caption{The left plot is model goodness of fit by a Latent Space Model, the right one is by cluster-specific ERGM using the membership estimates from the left model. Notes that although the estimation of parameters: (-0.87, 0.28, 0.20), (-5.00, 1.39, 0.58), (-3.84, 0.66, 1.38) are not close to the truth: (-2, 0.5, 0.5), (-2, 0.5, 0.5), (-2, 0.5, 0.5), the model fit is still very good.}
	\label{fig_gof}
\end{figure*}

One question a practitioner may ask is: if my working model is good enough, why should I bother to fit cluster specific ERGM? The answer is two folded. One is for estimation / hypothesis testing, the other is for overall goodness of fit. Figure \ref{fig_gof} shows that a second fine tuning step may greatly improve the model goodness of fit, even then the working model did a perfect job on clustering.
\subsubsection{Sensitivity to mis-clustering}
\subsection{A Degree Distribution Example}
Another major type of network dependence we would like to use as an example is the degree distribution. Since there is a specific spectral clustering method designed to the so called Degree Corrected Stochastic Block Models (DC-SBM) \cite{jin2015score}. We evaluate mis-clustering rate of that method on a degree distribution ERGM.

\section{Discussion and conclusion} \label{sec_disc}
In this paper, we analyze the complementary strengths and limitations of ERGM and LSM, both in model specification and the interpretation of parameters. We start from the computational non-scalability of the Bayesian inference approach for the Hierarchical ERGM and propose a two phase procedure as a feasible way to do data analysis. We intuitively formulate this procedure, that is to find clusters using Latent Space Models (LSM) first and then to fit cluster specific ERGM, conditioned on the first phase result. The key idea is to decouple the estimates of membership $M$ and ERGM parameters $\theta_k$, so we can provide a feasible way to improve goodness of fit, rather than to archive good asymptotic properties.\\
When modeling social networks or other types of relational data, valid statistical inference is especially challenging if only one single network is observed. This network can be viewed as a  snapshot of the accumulated effects of possibly more than one relation forming processes. So, our future direction along this line is to propose a new class of dynamic HERGM for longitudinal network data.

\bibliographystyle{abbrv}
\bibliography{}

\begin{thebibliography}{10}

\bibitem{besag1974spatial}
J.~Besag.
\newblock Spatial interaction and the statistical analysis of lattice systems.
\newblock {\em Journal of the Royal Statistical Society. Series B
  (Methodological)}, pages 192--236, 1974.

\bibitem{frank1986markov}
O.~Frank and D.~Strauss.
\newblock Markov graphs.
\newblock {\em Journal of the american Statistical association},
  81(395):832--842, 1986.

\bibitem{geman1984gibbs}
S.~Geman and D.~Geman.
\newblock Stochastic relaxation, gibbs distributions, and the bayesian
  restoration of images.
\newblock {\em Pattern Analysis and Machine Intelligence, IEEE Transactions
  on}, (6):721--741, 1984.

\bibitem{geyer1992constrained}
C.~J. Geyer and E.~A. Thompson.
\newblock Constrained monte carlo maximum likelihood for dependent data.
\newblock {\em Journal of the Royal Statistical Society. Series B
  (Methodological)}, pages 657--699, 1992.

\bibitem{gong1981twostage}
G.~Gong and F.~J. Samaniego.
\newblock Pseudo maximum likelihood estimation: theory and applications.
\newblock {\em The Annals of Statistics}, pages 861--869, 1981.

\bibitem{handcock2008statnet}
M.~S. Handcock, D.~R. Hunter, C.~T. Butts, S.~M. Goodreau, and M.~Morris.
\newblock statnet: Software tools for the representation, visualization,
  analysis and simulation of network data.
\newblock {\em Journal of statistical software}, 24(1):1548, 2008.

\bibitem{handcock2007cluster}
M.~S. Handcock, A.~E. Raftery, and J.~M. Tantrum.
\newblock Model-based clustering for social networks.
\newblock {\em Journal of the Royal Statistical Society: Series A (Statistics
  in Society)}, 170(2):301--354, 2007.

\bibitem{handcock2003assessing}
M.~S. Handcock, G.~Robins, T.~A. Snijders, J.~Moody, and J.~Besag.
\newblock Assessing degeneracy in statistical models of social networks.
\newblock Technical report, Citeseer, 2003.

\bibitem{hoff2002latent}
P.~D. Hoff, A.~E. Raftery, and M.~S. Handcock.
\newblock Latent space approaches to social network analysis.
\newblock {\em Journal of the american Statistical association},
  97(460):1090--1098, 2002.

\bibitem{hummel2012improving}
R.~M. Hummel, D.~R. Hunter, and M.~S. Handcock.
\newblock Improving simulation-based algorithms for fitting ergms.
\newblock {\em Journal of Computational and Graphical Statistics},
  21(4):920--939, 2012.

\bibitem{hunter2012goodness}
D.~R. Hunter, S.~M. Goodreau, and M.~S. Handcock.
\newblock Goodness of fit of social network models.
\newblock {\em Journal of the American Statistical Association}, 2012.

\bibitem{hunter2006curved}
D.~R. Hunter and M.~S. Handcock.
\newblock Inference in curved exponential family models for networks.
\newblock {\em Journal of Computational and Graphical Statistics}, 15(3), 2006.

\bibitem{ji2014coauthorship}
P.~Ji and J.~Jin.
\newblock Coauthorship and citation networks for statisticians.
\newblock {\em arXiv preprint arXiv:1410.2840}, 2014.

\bibitem{jin2015score}
J.~Jin.
\newblock Fast community detection by score.
\newblock {\em The Annals of Statistics}, 43(1):57--89, 2015.

\bibitem{krivitsky2008latentnet}
P.~N. Krivitsky and M.~Handcock.
\newblock Fitting position latent cluster models for social networks with
  latentnet.
\newblock {\em Journal of the Statistical Software}, 2008.

\bibitem{krivitsky2009representing}
P.~N. Krivitsky, M.~S. Handcock, A.~E. Raftery, and P.~D. Hoff.
\newblock Representing degree distributions, clustering, and homophily in
  social networks with latent cluster random effects models.
\newblock {\em Social networks}, 31(3):204--213, 2009.

\bibitem{lorrain1971block}
F.~Lorrain and H.~C. White.
\newblock Structural equivalence of individuals in social networks.
\newblock {\em The Journal of mathematical sociology}, 1(1):49--80, 1971.

\bibitem{meng2014trio}
X.-L. Meng.
\newblock A trio of inference problems that could win you a nobel prize in
  statistics (if you help fund it).
\newblock {\em Past, Present, and Future of Statistical Science}, 2014.

\bibitem{nowicki2001block}
K.~Nowicki and T.~A.~B. Snijders.
\newblock Estimation and prediction for stochastic blockstructures.
\newblock {\em Journal of the American Statistical Association},
  96(455):1077--1087, 2001.

\bibitem{pattison2002neighborhood}
P.~Pattison and G.~Robins.
\newblock Neighborhood-based models for social networks.
\newblock {\em Sociological Methodology}, 32(1):301--337, 2002.

\bibitem{schweinberger2015local}
M.~Schweinberger and M.~S. Handcock.
\newblock Local dependence in random graphs: characterization, properties, and
  statistical inference.
\newblock {\em Journal of the Royal Statistical Society: Series B (Statistical
  Methodology)}, 2015.

\bibitem{snijders2002markov}
T.~A. Snijders.
\newblock Markov chain monte carlo estimation of exponential random graph
  models.
\newblock {\em Journal of Social Structure}, 3(2):1--40, 2002.

\bibitem{snijders1997block}
T.~A. Snijders and K.~Nowicki.
\newblock Estimation and prediction for stochastic blockmodels for graphs with
  latent block structure.
\newblock {\em Journal of classification}, 14(1):75--100, 1997.

\bibitem{snijders2006new}
T.~A. Snijders, P.~E. Pattison, G.~L. Robins, and M.~S. Handcock.
\newblock New specifications for exponential random graph models.
\newblock {\em Sociological methodology}, 36(1):99--153, 2006.

\bibitem{strauss1990pseudolikelihood}
D.~Strauss and M.~Ikeda.
\newblock Pseudolikelihood estimation for social networks.
\newblock {\em Journal of the American Statistical Association},
  85(409):204--212, 1990.

\bibitem{tierney1994mixing}
L.~Tierney.
\newblock Markov chains for exploring posterior distributions.
\newblock {\em the Annals of Statistics}, pages 1701--1728, 1994.

\bibitem{van2007comparison}
M.~A. van Duijn, K.~Gile, and M.~S. Handcock.
\newblock Comparison of maximum pseudo likelihood and maximum likelihood
  estimation of exponential family random graph models.
\newblock Technical report, Working Paper 74, Center for Statistics and the
  Social Sciences, University of Washington. URL http://www. csss. washington.
  edu/Papers, 2007.

\bibitem{wasserman1996logit}
S.~Wasserman and P.~Pattison.
\newblock Logit models and logistic regressions for social networks: I. an
  introduction to markov graphs andp.
\newblock {\em Psychometrika}, 61(3):401--425, 1996.

\end{thebibliography}
\end{document}